\documentclass[aps,showpacs,preprintnumbers,amsmath, amssymb]{revtex4}

\usepackage{float}
\usepackage{graphics,epsfig}
\usepackage{graphicx}
\usepackage{dcolumn}
\usepackage{bm}

\begin{document}
\baselineskip=0.8 cm
\title{{\bf Holographic insulator/superconductor phase transition with Weyl corrections}}

\author{Zixu Zhao$^{1,2}$, Qiyuan Pan$^{1,2}$\footnote{panqiyuan@126.com} and Jiliang Jing$^{1,2}$\footnote{jljing@hunnu.edu.cn}}
\affiliation{$^{1}$Institute of Physics and Department of Physics,
Hunan Normal University, Changsha, Hunan 410081, China}
\affiliation{$^{2}$ Key Laboratory of Low Dimensional Quantum
Structures and Quantum Control of Ministry of Education, Hunan
Normal University, Changsha, Hunan 410081, China}

\vspace*{0.2cm}
\begin{abstract}
\baselineskip=0.6 cm
\begin{center}
{\bf Abstract}
\end{center}

We analytically investigate the phase transition between the holographic insulator and superconductor with Weyl corrections by using the variational method for the Sturm-Liouville eigenvalue problem. We find that similar to the curvature corrections, in p-wave model, the higher Weyl couplings make the insulator/superconductor phase transition harder to occur. However, in s-wave case the Weyl corrections do not influence the critical chemical potential, which is in contrast to the effect caused by the curvature corrections. Moreover, we observe that the Weyl corrections will not affect the critical phenomena and the critical exponent of the system always takes the mean-field value in both models. Our analytic results are found to be in good agreement with the numerical findings.

\end{abstract}
\maketitle
\newpage
\vspace*{0.2cm}

\section{Introduction}

The anti-de Sitter/conformal field theory (AdS/CFT) correspondence \cite{Maldacena}, which relates a string theory in AdS spacetime to a conformal field theory living on its boundary, has motivated the development of dual gravity models to describe strongly correlated systems in condensed matter physics \cite{GubserPRD78}. It was suggested that the gravitational duals to the high temperature superconducting systems consist of a black hole in an AdS spacetime with a complex scalar field coupled to a U(1) gauge field \cite{HartnollJHEP12}. When the temperature of the black hole is below a critical temperature $T_{c}$, the bulk configuration becomes unstable and experiences a second order phase transition, which exhibit the behavior of the superconductor in the boundary dual CFT \cite{HartnollPRL101}. Considering the potential applications to the condensed matter physics, many authors have constructed various s-, p- and d-wave holographic superconductor models in the AdS black hole background, for reviews, see Ref. \cite{SuperCondRev} and references therein.

Besides the bulk AdS black hole spacetime, the AdS soliton is a gravitational configuration which has lower
energy than the AdS space in the Poincar\'{e} coordinates, but has the same boundary topology as the Ricci flat black hole and the AdS space in the Poincar\'{e} coordinates \cite{HorowitzMyers}. Using a five-dimensional AdS soliton background coupled to a Maxwell field and a scalar field, Nishioka \emph{et al.} first constructed a model describing an insulator/superconductor phase transition at zero temperature in the probe limit where the backreaction of matter fields on the spacetime metric is neglected \cite{Nishioka-Ryu-Takayanagi}. It is found that when the chemical potential is sufficiently large beyond a critical value $\mu_{c}$, the AdS soliton becomes unstable to form scalar hair and a second order phase transition can happen, which can be used to describe the transition between the insulator and superconductor. Along this line, there have been accumulated interest to study various insulator and superconductor phase transitions in different theories of gravity \cite{Soliton,Akhavan-Soliton,Cai-Li-Zhang,PJWPRD,SLSoliton}.

In the previous papers \cite{Pan-Wang,PJWJHEP}, we extended the gravitational construction to include a Ricci flat AdS soliton in Gauss-Bonnet gravity \cite{Cai-Kim-Wang} and observed that the higher curvature corrections make it harder for the insulator/superconductor phase transition to occur. As a matter of fact, in order to understand the influences of the $1/N$ or $1/\lambda$ ($\lambda$ is the 't Hooft coupling) corrections on the holographic dual models, it is interesting to consider the higher derivative correction related to the gauge field besides the curvature correction to the gravity. Recently, Wu \emph{et al.} constructed an s-wave holographic dual model with Weyl corrections in order to explore the effects beyond the large $N$ limit on the holographic superconductor \cite{WuCKW}. They found that the higher Weyl corrections make it easier for the condensation to form, which is in strong contrast to the higher curvature corrections \cite{Gregory}. In the St$\ddot{u}$ckelberg mechanism, rich physics in the phase transition of the holographic superconductor with Weyl corrections has been observed \cite{MaCW}. More recently, the authors of \cite{MomeniSL} studied the p-wave holographic superconductor model with Weyl corrections and their results showed that the effect of Weyl corrections on the condensation is similar to that of the s-wave model. Considering that the increasing interest in investigation of Weyl corrections \cite{WuCKW,MaCW,MomeniSL,WeylC}, in this work we will consider the holographic insulator/superconductor phase transition model with Weyl corrections to the usual Maxwell field \cite{Ritz-Ward} in the probe limit, which has not been constructed as far as we know. Note that the variational method for the Sturm-Liouville (S-L) eigenvalue problem \cite{Gelfand-Fomin}, which was first developed by Siopsis and Therrien to analytically calculate the critical exponent near the critical temperature \cite{Siopsis}, is very effective to obtain the results on the condensation and the critical phenomena both in AdS black hole backgrounds \cite{SLAdSBH} and AdS soliton backgrounds \cite{PJWJHEP,Cai-Li-Zhang,PJWPRD,SLSoliton}. Thus, we will generalize the S-L method to study holographic insulator/superconductor phase transition with Weyl corrections in the AdS soliton background. It is not trivial to analytically study the condensation and the phase transition by taking into account of the influence of the Weyl couplings. Besides to be used to check numerical computation, the analytic investigation can clearly present the critical exponent of the system at the critical point and the influence of the Weyl correction terms on the phase transition.

The plan of the work is the following. In Sec. II we explore the p-wave insulator/superconductor phase transition with Weyl corrections. In particular, we calculated the critical chemical potential of the system as well as the relations of condensed values of operators and the charge density with respect to $(\mu-\mu_{c})$. In Sec. III we discuss the s-wave case. We conclude in the last section with our main results.

\section{P-wave insulator/superconductor phase transition with Weyl corrections}

In this section, we will study the model of the p-wave insulator/superconductor phase transition with Weyl corrections in the five-dimensional AdS soliton spacetime by considering an $SU(2)$ Yang-Mills action with Weyl corrections in the bulk theory \cite{MomeniSL}
\begin{eqnarray}\label{p-System}
S=\int d^{5}x\sqrt{-g}\left[\frac{1}{2\kappa^2}\left(R+\frac{12}{L^2}\right)
-\frac{1}{4\hat{g}^2}\left(F^{a}_{\mu\nu}F^{a\mu\nu}-4\gamma
C^{\mu\nu\rho\sigma}F^{a}_{\mu\nu}F^{a}_{\rho\sigma}\right)\right],
\end{eqnarray}
where $\kappa^{2}=8\pi G_{5}$ is the five-dimensional gravitational constant, $L$ is the AdS radius, $\hat{g}$ is the Yang-Mills coupling constant, and $\gamma$ is the so-called Weyl coupling parameter which satisfies $-L^{2}/16<\gamma<L^{2}/24$ \cite{Ritz-Ward}. $F^{a}_{\mu\nu}=\partial_{\mu}A^{a}_{\nu}-\partial_{\nu}A^{a}_{\mu}+\epsilon^{abc}A^{b}_{\mu}A^{c}_{\nu}$
is the $SU(2)$ Yang-Mills field strength and $\epsilon^{abc}$ is the totally antisymmetric tensor with $\epsilon^{123}=+1$. The $A^{a}_{\mu}$ are the components of the mixed-valued gauge fields $A=A^{a}_{\mu}\tau^{a}dx^{\mu}$, where $\tau^{a}$ are the three generators of the $SU(2)$ algebra with commutation relation $[\tau^{a},\tau^{b}]=\epsilon^{abc}\tau^{c}$.

In this Letter, we will construct the model of holographic insulator/superconductor phase transition in the probe limit where the backreaction of matter fields on the metric can be neglected. Due to the scaling symmetries of the system for the case of the p-wave \cite{PWave}, we can see from the action (\ref{p-System}) that the probe limit can be obtained safely if the coupling constant $\hat{g}$ is large enough, i.e., $\kappa^2/\hat{g}^2\rightarrow0$. Without loss of generality, we will set $\hat{g}=1$ and work in this probe approximation.

In the probe limit, the background metric is a five-dimensional AdS soliton
\begin{eqnarray}\label{soliton}
ds^2=-r^{2}dt^{2}+\frac{dr^2}{f(r)}+f(r)d\varphi^2+r^{2}(dx^{2}+dy^{2}),
\end{eqnarray}
with
\begin{eqnarray}
f(r)=\frac{r^2}{L^{2}}(1-\frac{r_{s}^{4}}{r^{4}}),
\end{eqnarray}
where $r_{s}$ is the tip of the soliton which is a conical
singularity in this solution. It should be noted that we can remove the singularity by imposing a period $\beta=\pi L^{2}/r_{s}$ for the coordinate $\varphi$. In fact, this soliton can be obtained from a five-dimensional AdS Schwarzschild black hole by making use of two Wick rotations. Just as in Ref. \cite{Nishioka-Ryu-Takayanagi}, we will take the AdS radius $L=1$ in our discussion for clarity. Since we are interested in the Weyl corrections to the holographic insulator/superconductor phase transition, we will use the nonzero components of the Weyl tensor $C_{\mu\nu\rho\sigma}$ for this considered solution
\begin{eqnarray}\label{WeylTensor}
&&C_{0i0j}=-r_s^4\delta_{ij},~~C_{0r0r}=\frac{r_s^4}{r^4-r_s^4},~~
C_{0\varphi0\varphi}=r_s^4\left(1-\frac{r_s^4}{r^4}\right),~~C_{r\varphi r\varphi}=\frac{3r_s^4}{r^4},\nonumber\\
&&C_{irjr}=-\frac{r_s^4}{r^4-r_s^4}\delta_{ij},~~C_{i\varphi j\varphi}=-r_s^4\left(1-\frac{r_s^4}{r^4}\right)\delta_{ij},~~
C_{ijkl}=r_s^4\delta_{ik}\delta_{jl},
\end{eqnarray}
with $i,~j,~k,~l=x$ or $y$.

In order to construct a p-wave holographic insulator and
superconductor, we adopt the ansatz of the gauge fields as \cite{Akhavan-Soliton,Cai-Li-Zhang,PWave}
\begin{eqnarray}\label{p-ansatz}
A(r)=\phi(r)\tau^{3}dt+\psi(r)\tau^{1}dx,
\end{eqnarray}
where we regard the $U(1)$ symmetry generated by $\tau^{3}$ as the
$U(1)$ subgroup of $SU(2)$ and the gauge boson with nonzero component
$\psi(r)$ along $x$-direction is charged under $A^{3}_{t}=\phi(r)$.
According to AdS/CFT correspondence, $\psi(r)$ is dual to the $x$-component of some charged
vector operator $\cal O$ on the boundary and $\phi(r)$ is dual to the chemical potential. The
condensation of $\psi(r)$ will spontaneously break the $U(1)$ gauge
symmetry and induce a phase transition, which can be interpreted as
a p-wave insulator and superconductor phase transition in the boundary field theory.

From the Yang-Mills action with Weyl corrections (\ref{p-System}),
we have the following equations of motion
\begin{eqnarray}
\left(1+\frac{8\gamma r_{s}^{4}}{r^4}\right)\psi^{\prime\prime}
+\left[\frac{1}{r}\left(1-\frac{24\gamma r_{s}^{4}}{r^4}\right)+\frac{f^\prime}{f}\left(1+\frac{8\gamma r_{s}^{4}}{r^4}\right)\right]\psi^\prime
+\left(1-\frac{8\gamma r_s^4}{r^4}\right)\frac{\phi^2}{r^2f}\psi=0,
\label{pwave-Psi}
\end{eqnarray}
\begin{eqnarray}
\left(1+\frac{8\gamma r_{s}^{4}}{r^4}\right)\phi^{\prime\prime}
+\left[\frac{1}{r}\left(1-\frac{24\gamma r_{s}^{4}}{r^4}\right)+\frac{f^\prime}{f}\left(1+\frac{8\gamma r_{s}^{4}}{r^4}\right)\right]\phi^\prime-
\left(1-\frac{8\gamma r_s^4}{r^4}\right)\frac{\psi^2}{r^2f}\phi=0,
\label{pwave-Phi}
\end{eqnarray}
where the prime denotes the derivative with respect to $r$.

In order to solve the above equations of motion, we have to impose
the appropriate boundary conditions for $\phi(r)$ and $\psi(r)$ at
the tip $r=r_{s}$ and at the boundary $r\rightarrow\infty$. The boundary conditions at the tip $r=r_{s}$ are
\begin{eqnarray}\label{PWBoundary}
&&\psi=\alpha_{0}+\alpha_{1}(r-r_{s})+\alpha_{2}(r-r_{s})^{2}+\cdots\,, \nonumber \\
&&\phi=\beta_{0}+\beta_{1}(r-r_{s})+\beta_{2}(r-r_{s})^{2}+\cdots\,,
\end{eqnarray}
where $\alpha_{i}$ and $\beta_{i}$ ($i=0,1,2,\cdots$) are the integration constants, and we have imposed the Neumann-like boundary conditions to render the physical quantities finite \cite{Nishioka-Ryu-Takayanagi}. Obviously, we can find a constant nonzero gauge field $\phi(r_{s})$ at $r=r_{s}$, which is in strong
contrast to that of the AdS black hole where $\phi(r_{+})=0$ at the
horizon.

At the asymptotic AdS boundary $r\rightarrow\infty$, the solutions behave like
\begin{eqnarray}
\psi=\psi_{0}+\frac{\psi_{2}}{r^{2}}\,,\hspace{0.5cm}
\phi=\mu-\frac{\rho}{r^{2}}\,, \label{PWinfinity}
\end{eqnarray}
where $\psi_{0}$ and $\psi_{2}$ may be identified as a source and the expectation value
of the dual operator, while $\mu$ and $\rho$ are interpreted as the chemical potential and
charge density in the boundary field theory, respectively. Making use the asymptotic condition $\psi_{0}=0$, we can obtain a normalizable solution since we are interested in the case where the dual operator is not sourced. For simplicity, we will scale $r_{s}=1$ in the following just as in \cite{Nishioka-Ryu-Takayanagi}.

\subsection{Critical chemical potential}

Let us change the coordinate and set $z=1/r$. Under this transformation, we can express the equations of
motion (\ref{pwave-Psi}) and (\ref{pwave-Phi}) as
\begin{eqnarray}
(1+8\gamma z^{4})\psi^{\prime\prime}
+\left[(1+40\gamma z^{4})\frac{1}{z}+(1+8\gamma z^{4})\frac{f^\prime}{f}\right]\psi^\prime
+(1-8\gamma z^4)\frac{\phi^2}{z^2f}\psi=0,
\label{PWPsiz}
\end{eqnarray}
\begin{eqnarray}
(1+8\gamma z^{4})\phi^{\prime\prime}
+\left[(1+40\gamma z^{4})\frac{1}{z}+(1+8\gamma z^{4})\frac{f^\prime}{f}\right]\phi^\prime-
(1-8\gamma z^4)\frac{\psi^2}{z^2f}\phi=0,
\label{PWPhiz}
\end{eqnarray}
where the prime now denotes the derivative with respect to $z$.

It is well-known that, adding the chemical potential to the
AdS soliton, the solution is unstable to develop a hair for the chemical potential
bigger than a critical value, i.e., $\mu>\mu_{c}$. On the
other hand, for lower chemical potential $\mu<\mu_{c}$, the scalar field is zero and it
can be interpreted as the insulator phase since in this model the normal phase is described
by an AdS soliton where the system exhibits mass gap \cite{Nishioka-Ryu-Takayanagi}. Thus, the turning point of the holographic insulator/superconductor phase transition is the critical chemical potential $\mu_{c}$.

\begin{figure}[ht]
\includegraphics[scale=0.772]{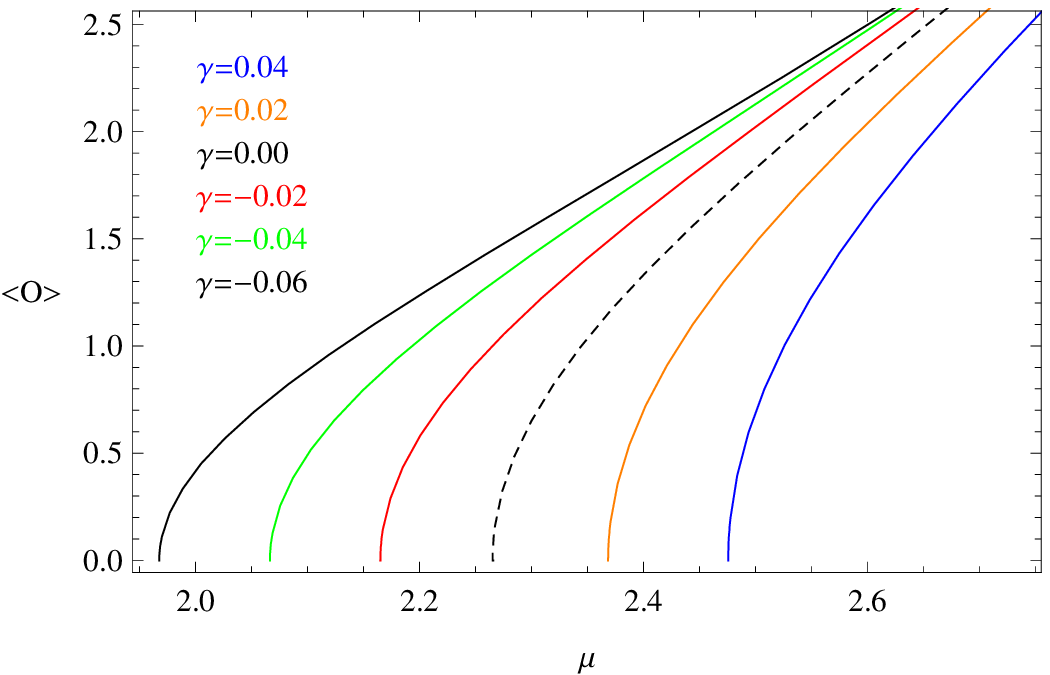}\vspace{0.0cm}
\includegraphics[scale=0.75]{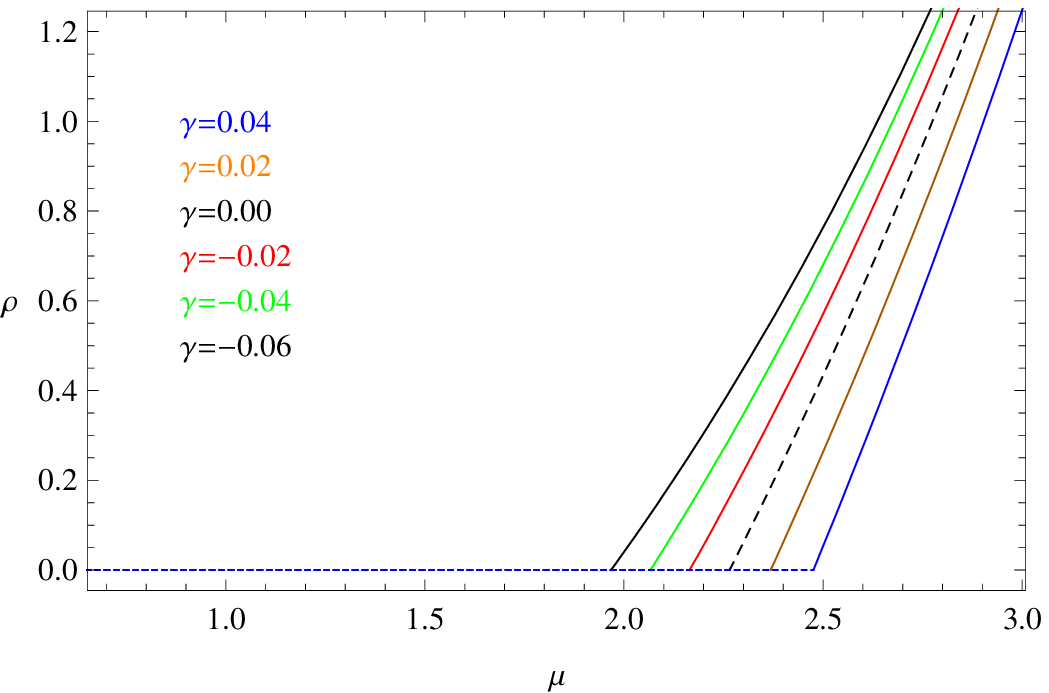}\\ \vspace{0.0cm}
\caption{\label{PWaveCondRL} (color online) The condensate of the operator $\langle{\cal O}\rangle=\psi_{2}$ and charge density $\rho$ with respect to the chemical potential $\mu$ for different Weyl coupling parameters $\gamma$ for the p-wave holographic insulator
and superconductor model. In each panel, the six lines from left to right
correspond to increasing $\gamma$, i.e., $\gamma=-0.06$ (black),
$-0.04$ (green), $-0.02$ (red), $0$ (black and dashed), $0.02$ (orange) and
$0.04$ (blue) respectively.}
\end{figure}

Since the scalar field $\psi=0$ at the critical chemical potential $\mu_{c}$, so below the critical point Eq. (\ref{PWPhiz}) reduces to
\begin{eqnarray}
(1+8\gamma z^{4})\phi^{\prime\prime}
+\left[(1+40\gamma z^{4})\frac{1}{z}+(1+8\gamma z^{4})\frac{f^\prime}{f}\right]\phi^\prime=0.
\label{PWPhiCritical}
\end{eqnarray}
Considering the Neumann-like boundary condition (\ref{PWBoundary}) for the gauge field $\phi$ at the tip $z=1$, we can get the physical solution $\phi(z)=\mu$ to Eq. (\ref{PWPhiCritical}) if $\mu<\mu_{c}$, which indicates that close to the critical point $\rho=0$ according to the asymptotic behavior in Eq. (\ref{PWinfinity}) near the AdS boundary $z=0$. This is consistent with the numerical results in Fig. \ref{PWaveCondRL}
which plots the condensate of the operator $\langle{\cal O}\rangle=\psi_{2}$ and charge density $\rho$ with respect to the chemical potential $\mu$ for different Weyl coupling parameters $\gamma$, where $\rho=0$ when $\mu<\mu_{c}$.

As $\mu\rightarrow\mu_{c}$ from below the critical point, Eq. (\ref{PWPsiz}) will become
\begin{eqnarray}
(1+8\gamma z^{4})\psi^{\prime\prime}
+\left[(1+40\gamma z^{4})\frac{1}{z}+(1+8\gamma z^{4})\frac{f^\prime}{f}\right]\psi^\prime
+(1-8\gamma z^4)\frac{\mu^2}{z^2f}\psi=0,
\label{NewPWPsiCritical}
\end{eqnarray}
which is the master equation to give the critical chemical potential $\mu_{c}$ in the S-L method. Introducing a trial function $F(z)$ near the boundary $z=0$
just as in \cite{Siopsis}
\begin{eqnarray}\label{PWPhiFz}
\psi(z)\sim \langle{\cal O}\rangle z^2F(z),
\end{eqnarray}
with the boundary condition $F(0)=1$ and $F'(0)=0$, we can obtain the equation of motion for $F(z)$ \begin{eqnarray}\label{NewPWFzmotion}
(TF^{\prime})^{\prime}+(U+\mu^2 V)F=0,
\end{eqnarray}
where
\begin{eqnarray}
T(z)=z^{3}(z^{4}-1)(1+8\gamma z^4),~~U=2\left[2z(z^{4}-1)(1+24\gamma z^{4})+\frac{Tf^\prime}{zf}\right],
~~V=\frac{z(z^{4}-1)(1-8\gamma z^4)}{f}.
\end{eqnarray}
We find that, according to the Sturm-Liouville eigenvalue problem
\cite{Gelfand-Fomin}, the minimum eigenvalue of $\mu^2$ can be
obtained from variation of the following functional
\begin{eqnarray}\label{PWEigenvalue}
&\mu^{2}&=\frac{\int^{1}_{0}(TF'^{2}-UF^{2})dz}{\int^{1}_{0}VF^2dz}
\nonumber\\&&
=\frac{16[14(5+8\gamma)-35(3+8\gamma)a+8(7+24\gamma)a^{2}]}
{7[30(1-4\gamma)-8(5-24\gamma)a+5(3-16\gamma)a^{2}]},
\end{eqnarray}
where we have used the trial function $F(z)=1-az^{2}$ with a constant $a$. For different Weyl coupling parameters $\gamma$, we can get the minimum eigenvalues of $\mu^{2}$
and the corresponding values of $a$, for example, $\mu_{min}^{2}=3.888$ and $a=0.0264$ for
$\gamma=-0.06$, $\mu_{min}^{2}=4.280$ and $a=0.159$ for
$\gamma=-0.04$, $\mu_{min}^{2}=4.695$ and $a=0.259$ for
$\gamma=-0.02$, $\mu_{min}^{2}=5.139$ and $a=0.338$ for
$\gamma=0$, $\mu_{min}^{2}=5.621$ and $a=0.403$ for
$\gamma=0.02$, and $\mu_{min}^{2}=6.147$ and $a=0.458$ for
$\gamma=0.04$. Then, we have the critical chemical potential $\mu_{c}=\mu_{min}$ \cite{Cai-Li-Zhang}.
In Table \ref{PWCritical}, we give the critical chemical potential $\mu_{c}$ for chosen values of Weyl coupling parameters. In order to compare with numerical
results, we also present the critical chemical potential $\mu_{c}$ obtained by
using the shooting method. Obviously, one can find that the analytic results derived from S-L method are in good agreement with the numerical calculation.

\begin{table}[ht]
\begin{center}
\caption{\label{PWCritical} The critical chemical potential $\mu_{c}$ obtained by the analytical S-L method and from numerical calculation with fixed Weyl coupling parameters for the p-wave holographic insulator and superconductor model. Note that our result reduces to the result in Refs.
\cite{Akhavan-Soliton,Cai-Li-Zhang} if $\gamma\rightarrow0$.}
\begin{tabular}{c c c c c c c}
         \hline \hline
$\gamma$ & -0.06 & -0.04 & -0.02 & 0 & 0.02 & 0.04
        \\
        \hline
~~~~$Analytical$~~~~~&~~~~~$1.972$~~~~~&~~~~~$2.069$~~~~~&~~~~~$2.167$~~~~~&~~~~~$2.267$~~~~~&~~~~~$2.371$~~~~~&~~~~~$2.479$~~~~
          \\
~~~~$Numerical$~~~~~&~~~~~$1.968$~~~~~&~~~~~$2.066$~~~~~&~~~~~$2.165$~~~~~&~~~~~$2.265$~~~~~&~~~~~$2.368$~~~~~&~~~~~$2.476$~~~~
          \\
        \hline \hline
\end{tabular}
\end{center}
\end{table}

From Table \ref{PWCritical}, we also find that the critical
chemical potential $\mu_{c}$ increases as we amplify the Weyl coupling
parameter $\gamma$, which shows that the higher Weyl corrections in general will make it harder for the phase transition between holographic insulator and superconductor to be triggered, just as the influences of the Gauss-Bonnet corrections on the p-wave holographic insulator/superconductor phase transition \cite{PJWJHEP}. This property agrees well with the numerical finding shown in
Fig. \ref{PWaveCondRL}.

\subsection{Critical phenomena}

We will use the S-L method to deal with the effect of the Weyl corrections on the critical phenomena for the phase transition between the p-wave holographic insulator and superconductor, especially the critical exponent for condensation operator and the relation between the charge density and the chemical potential.

When $\mu\rightarrow\mu_{c}$, the condensation of the operator $\langle{\cal O}\rangle$ is very small, we can expand $\phi(z)$ in $\langle{\cal O}\rangle$ as
\begin{eqnarray}\label{PWPhiChiz}
\phi(z)\sim\mu_{c}+\langle{\cal O}\rangle\chi(z)+\cdots,
\end{eqnarray}
where we have introduced the boundary condition $\chi(1)=0$ at the tip. It should be noted that Eq. (\ref{PWPhiChiz}) is true near the critical point $\mu_{c}$, but only above $\mu_{c}$. Substituting the
function (\ref{PWPhiFz}) and (\ref{PWPhiChiz}) into (\ref{PWPhiz}), one can get the equation of
motion for $\chi(z)$
\begin{eqnarray}\label{PWChiz}
(Q\chi^{\prime})^{\prime}-\langle{\cal O}\rangle\mu_{c}(z^{4}-1)(1-8\gamma z^4)\frac{zF^2}{f}=0,
\end{eqnarray}
with a new function
\begin{eqnarray}\label{PWQz}
Q(z)=\frac{(z^{4}-1)(1+8\gamma z^{4})}{z}.
\end{eqnarray}
Obviously, the general solution for above equation takes the
form
\begin{eqnarray}\label{PWGSChiz}
&\chi(z)&=\langle{\cal O}\rangle\mu_{c}\xi(z)
\nonumber\\
&&=\langle{\cal O}\rangle\mu_{c}\left\{c_{1}+\int_{1}^{z}\left[c_{2}+\int_{1}^{y}(x^{4}-1)(1-8\gamma x^4)\frac{xF(x)^2}{f(x)}dx\right]\frac{1}{Q(y)}dy\right\},
\end{eqnarray}
where $c_{1}$ and $c_{2}$ are the integration constants which can be determined by the
boundary condition of $\chi(z)$.

Near the boundary $z=0$, we can also expand $\phi$ as
\begin{eqnarray}\label{PWEXPPhi}
\phi(z)\simeq\mu-\rho z^{2}\simeq\mu_{c}+\langle{\cal
O}\rangle[\chi(0)+\chi'(0)z+\frac{1}{2}\chi''(0)z^{2}+\cdots].
\end{eqnarray}
From the coefficients of the $z^{0}$ term in both sides of the
above formula, we can obtain
\begin{eqnarray}
\mu-\mu_{c}\simeq\langle{\cal O}\rangle\chi(0),
\end{eqnarray}
which gives
\begin{eqnarray}\label{PWOperator}
\langle{\cal O}\rangle=\frac{1}{[\mu_{c}\xi(0)]^{1/2}}(\mu-\mu_{c})^{1/2},
\end{eqnarray}
where $\xi(0)$ can be calculated via Eq. (\ref{PWGSChiz}). For example, we can get
$\langle{\cal O}\rangle\approx1.607(\mu-\mu_{c})^{1/2}=2.256(\frac{\mu}{\mu_{c}}-1)^{1/2}$ for $\gamma=-0.06$ when $a=0.0264$, which is in good agreement with the numerical result given in Fig. \ref{PWaveCondRL}. It should be noted that for the case of $\gamma=0$ when $a=0.338$, we obtain $\langle{\cal O}\rangle\approx2.560(\mu-\mu_{c})^{1/2}=3.855(\frac{\mu}{\mu_{c}}-1)^{1/2}$, which agrees with the result given in \cite{Akhavan-Soliton,Cai-Li-Zhang}.

Notice that the relation (\ref{PWOperator}) is valid
for all cases considered here, so the condensation $\langle{\cal
O}\rangle\sim(\mu-\mu_{c})^{1/2}$ near the critical point
for various values of Weyl coupling parameters $\gamma$, which is consistent with the
numerical results shown in Fig. \ref{PWaveCondRL} that the phase transition between the p-wave holographic insulator and superconductor with Weyl corrections belongs to the second order and the critical exponent
of the system takes the mean-field value $1/2$.

Comparing the coefficients of the $z^{1}$ term in Eq.
(\ref{PWEXPPhi}), we find that $\chi'(0)\rightarrow0$ which agrees well with the following relation by making integration of both sides of Eq. (\ref{PWChiz})
\begin{eqnarray}\label{PWChiz0}
\left[\frac{\chi'(z)}{z}\right]\bigg|_{z\rightarrow 0}=\langle{\cal
O}\rangle\mu_{c}\int_{0}^{1}(z^{4}-1)(1-8\gamma z^4)\frac{zF^2}{f}dz.
\end{eqnarray}

Considering the coefficients of the $z^{2}$ term in Eq. (\ref{PWEXPPhi}),
we arrive at
\begin{eqnarray}
\rho=-\frac{1}{2}\langle{\cal
O}\rangle\chi''(0)=\Gamma(\gamma)(\mu-\mu_{c}),
\end{eqnarray}
where $\Gamma(\gamma)$ is only the function of the Weyl coupling parameter $\gamma$ which can be given by
\begin{eqnarray}
\Gamma(\gamma)=-\frac{1}{2\xi(0)}\int_{0}^{1}(z^{4}-1)(1-8\gamma z^4)\frac{zF^2}{f}dz.
\end{eqnarray}
As an example, we calculate the case for $\gamma=-0.06$ and obtain $\Gamma(\gamma)=0.760$ when $a=0.0264$, i.e., the linear relation $\rho=0.760(\mu-\mu_{c})$,
which agrees well with the result shown in Fig. \ref{PWaveCondRL}. For the case of $\gamma=0$ when $a=0.338$, we have $\Gamma(\gamma)=1.126$ which results in $\rho=1.126(\mu-\mu_{c})$, just as presented in \cite{Akhavan-Soliton,Cai-Li-Zhang}. Here we notice that the Weyl corrections will not change the linear relation between the charge density and the chemical potential $\rho\sim(\mu-\mu_{c})$, which is again in good agreement with the numerical result plotted in Fig. \ref{PWaveCondRL}.

\section{S-wave insulator/superconductor phase transition with Weyl corrections}

In order to construct an s-wave holographic insulator and superconductor with Weyl corrections in the AdS soliton spacetime, we will consider a Maxwell field and a charged complex scalar field coupled via the action  \cite{WuCKW,MaCW}
\begin{eqnarray}\label{SWAction}
S=\int
d^{5}x\sqrt{-g}\left[\frac{1}{2\kappa^2}\left(R+\frac{12}{L^2}\right)-\frac{1}{4}\left(F_{\mu\nu}F^{\mu\nu}-4\gamma
C^{\mu\nu\rho\sigma}F_{\mu\nu}F_{\rho\sigma}\right)-|\nabla\psi -
iqA\psi|^{2} -m^2|\psi|^2\right],
\end{eqnarray}
where $A$ is the gauge field, $F_{\mu\nu}=\partial_{\mu}A_{\nu}-\partial_{\nu}A_{\mu}$ is the field strength tensor, $q$ and $m$ represent the charge and mass of the scalar field $\psi$ respectively. We will be working in the probe approximation, which is equivalent to letting $q\rightarrow\infty$ by using the scaling symmetries of the s-wave system, i.e., $\kappa^2/q^2\rightarrow0$. Without loss of generality, we can set $q=1$ just as in Refs. \cite{WuCKW,Nishioka-Ryu-Takayanagi}.

Taking the ansatz of the matter fields as $\psi=\psi(r)$ and
$A=\phi(r) dt$, we can obtain the equations of motion from the action (\ref{SWAction}) for the
scalar field $\psi$ and gauge field $\phi$ in the probe limit
\begin{eqnarray}
\psi^{\prime\prime}+\left(
\frac{3}{r}+\frac{f^\prime}{f}\right)\psi^\prime
+\left(\frac{\phi^2}{r^2f}-\frac{m^2}{f}\right)\psi=0\,, \label{SWPsi}
\end{eqnarray}
\begin{eqnarray}
\left(1+\frac{8\gamma r_{s}^{4}}{r^4}\right)\phi^{\prime\prime}
+\left[\frac{1}{r}\left(1-\frac{24\gamma r_{s}^{4}}{r^4}\right)+\frac{f^\prime}{f}\left(1+\frac{8\gamma r_{s}^{4}}{r^4}\right)\right]\phi^\prime-\frac{2\psi^2}{f}\phi=0,
\label{SWPhi}
\end{eqnarray}
where the prime denotes the derivative with respect to $r$.

Considering the boundary conditions at the tip $r=r_{s}$, we find that the solutions have the same form just as Eq. (\ref{PWBoundary}) for the p-wave holographic insulator and superconductor model with Weyl corrections. But near the boundary $r\rightarrow\infty$, we get different asymptotic
behaviors
\begin{eqnarray}
\psi=\frac{\psi_{-}}{r^{\Delta_{-}}}+\frac{\psi_{+}}{r^{\Delta_{+}}}\,,\hspace{0.5cm}
\phi=\mu-\frac{\rho}{r^{2}}\,, \label{SWInfinity}
\end{eqnarray}
with $\Delta_\pm=2\pm\sqrt{4+m^{2}}$. Provided $\Delta_{-}$ is larger than the unitarity
bound, the coefficients $\psi_{-}$ and $\psi_{+}$ both multiply normalizable modes of the scalar field equations and they correspond to the vacuum expectation values $\psi_{-}=\langle{\cal O}_{-}\rangle$, $\psi_{+}=\langle{\cal O}_{+}\rangle$ of operators dual to the scalar field according to the AdS/CFT correspondence. We can impose boundary conditions that either $\psi_{-}$ or $\psi_{+}$ vanish \cite{HartnollJHEP12,HartnollPRL101}.

\subsection{Critical chemical potential}

Introducing the variable $z=1/r$, we can convert the equations of motion (\ref{SWPsi}) and (\ref{SWPhi}) to be
\begin{eqnarray}
\psi^{\prime\prime}+\left(
\frac{f^\prime}{f}-\frac{1}{z}\right)\psi^\prime
+\left(\frac{\phi^2}{z^2f}-\frac{m^2}{z^4f}\right)\psi=0\,,
\label{SWPsiz}
\end{eqnarray}
\begin{eqnarray}
(1+8\gamma z^{4})\phi^{\prime\prime}+\left[(1+40\gamma z^{4})\frac{1}{z}+(1+8\gamma z^{4})\frac{f^\prime}{f}\right]\phi^\prime-\frac{2\psi^2}{z^4f}\phi=0. \label{SWPhiz}
\end{eqnarray}
Here the prime denotes the derivative with respect to $z$.

\begin{figure}[ht]
\includegraphics[scale=0.8]{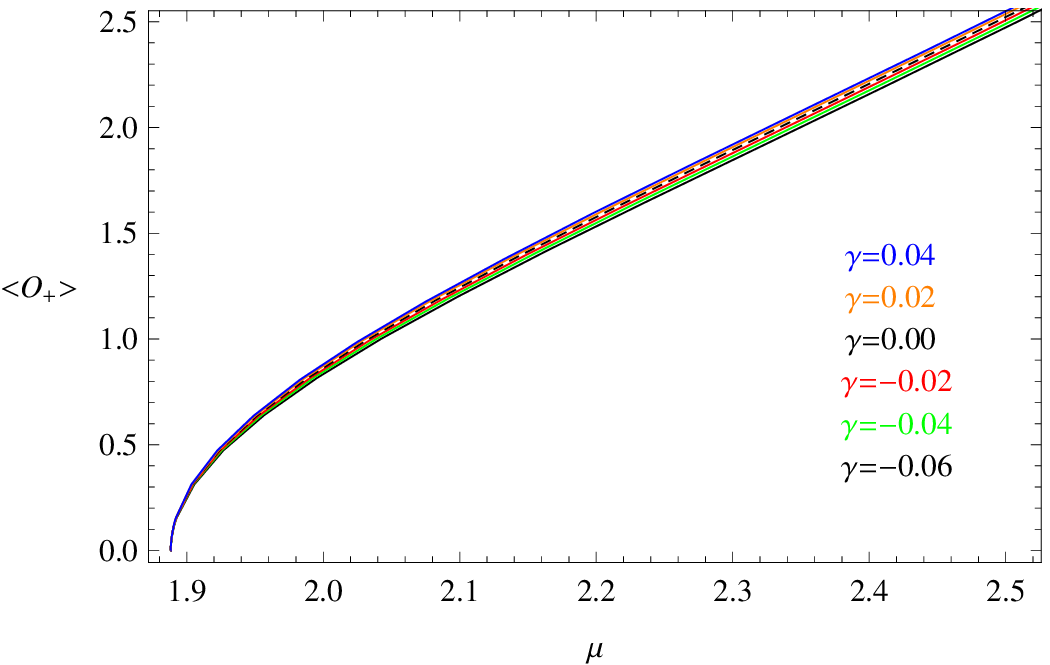}\vspace{0.0cm}
\includegraphics[scale=0.75]{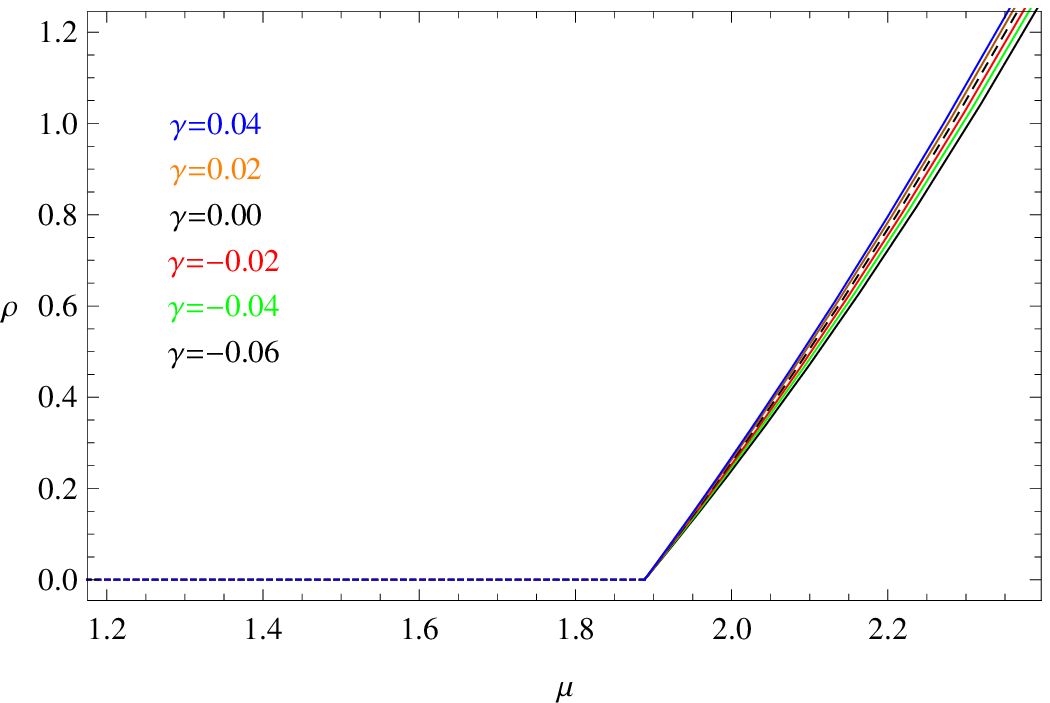}\\ \vspace{0.0cm}
\caption{\label{SWaveCondRLZ} (color online) The condensate of the operator $\langle{\cal O_{+}}\rangle$ and charge density $\rho$ with respect to the chemical potential $\mu$ for different Weyl coupling parameters $\gamma$ with the mass of the scalar field $m^{2}L^{2}=-15/4$ for the s-wave holographic insulator and superconductor model. In each panel, the six lines from bottom to top correspond to increasing $\gamma$, i.e., $\gamma=-0.06$ (black), $-0.04$ (green), $-0.02$ (red), $0$ (black and dashed), $0.02$ (orange) and $0.04$ (blue) respectively.}
\end{figure}

\begin{figure}[ht]
\includegraphics[scale=0.8]{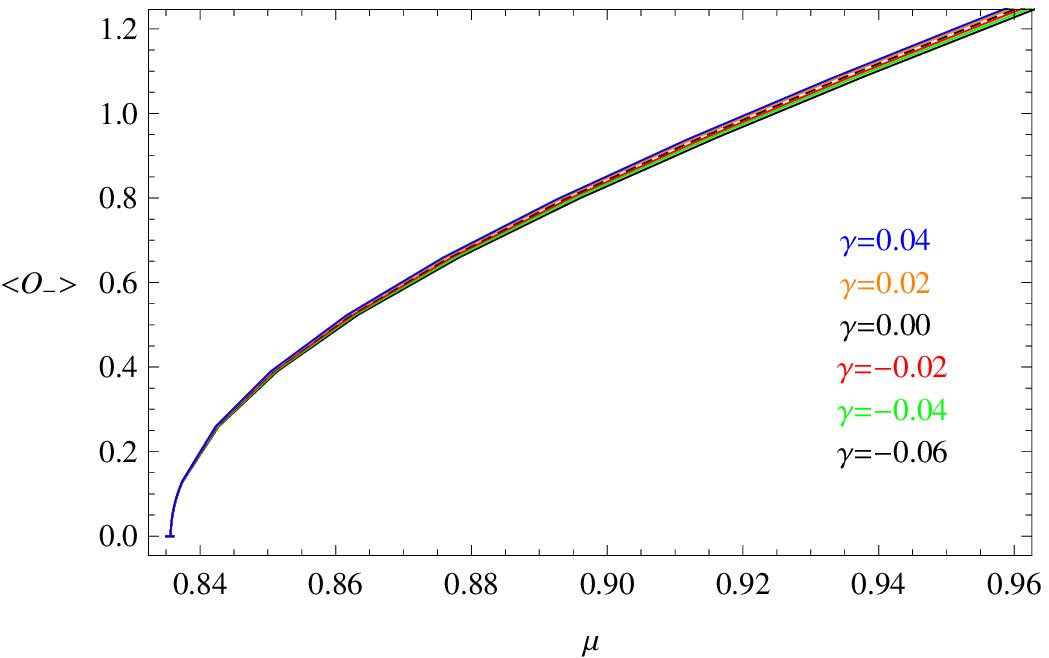}\vspace{0.0cm}
\includegraphics[scale=0.75]{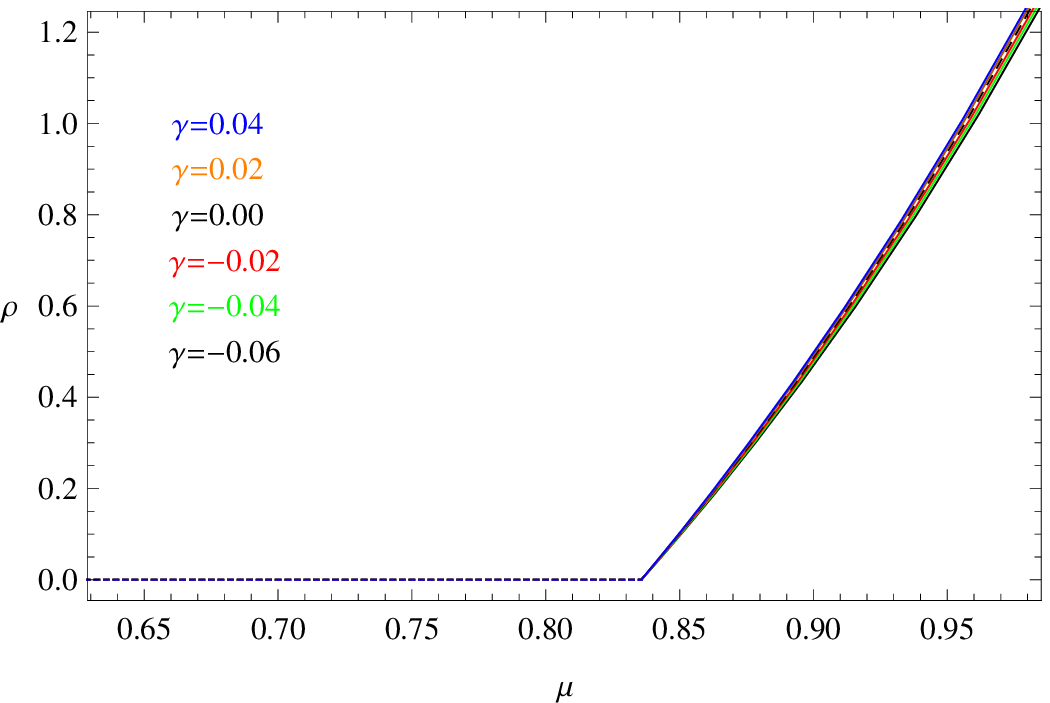}\\ \vspace{0.0cm}
\caption{\label{SWaveCondRLF} (color online) The condensate of the operator $\langle{\cal O_{-}}\rangle$ and charge density $\rho$ with respect to the chemical potential $\mu$ for different Weyl coupling parameters $\gamma$ with the mass of the scalar field $m^{2}L^{2}=-15/4$ for the s-wave holographic insulator and superconductor model. In each panel, the six lines from bottom to top correspond to increasing $\gamma$, i.e., $\gamma=-0.06$ (black), $-0.04$ (green), $-0.02$ (red), $0$ (black and dashed), $0.02$ (orange) and $0.04$ (blue) respectively.}
\end{figure}

At the critical chemical potential $\mu_{c}$, the scalar field
$\psi=0$. Thus, below the critical point Eq. (\ref{SWPhiz}) reduces to
\begin{eqnarray}
(1+8\gamma z^{4})\phi^{\prime\prime}+\left[(1+40\gamma z^{4})\frac{1}{z}+(1+8\gamma z^{4})\frac{f^\prime}{f}\right]\phi^\prime=0.
\label{SWPhiCritical}
\end{eqnarray}
Similar to the analysis in the previous section, we can obtain the physical solution $\phi(z)=\mu$ to Eq. (\ref{SWPhiCritical}) when $\mu<\mu_{c}$. Note that the asymptotic behavior in Eq.
(\ref{SWInfinity}), close to the critical point $\mu_{c}$, this solution implies that $\rho=0$ near the AdS boundary $z=0$, which is in good agreement with numerical findings obtained from Figs. \ref{SWaveCondRLZ} and \ref{SWaveCondRLF} where we plot the condensate of the operator $\langle{\cal O}_{i}\rangle$ ($i=+$ or $i=-$) and charge density $\rho$ with respect to the chemical potential $\mu$ for different Weyl coupling parameters $\gamma$.

As $\mu\rightarrow\mu_{c}$ from below the critical point, the scalar field equation (\ref{SWPsiz}) becomes
\begin{eqnarray}
\psi^{\prime\prime}+\left(\frac{f^\prime}{f}-\frac{1}{z}\right)\psi^\prime
+\left(\frac{\mu^2}{z^2f}-\frac{m^2}{z^4f}\right)\psi=0,
\label{SWCriticalPsi}
\end{eqnarray}
which is the master equation to calculate the critical chemical potential
$\mu_{c}$ in the S-L method. It should be noted that, although Eq. (\ref{SWPhiCritical}) for the gauge field $\phi$ depends on $\gamma$, but the Weyl coupling parameters $\gamma$ are absent in the master Eq. (\ref{SWCriticalPsi}), which leads that the Weyl corrections do not have any effect on the critical chemical potential $\mu_{c}$ for the fixed mass of the scalar field, just as shown in Figs. \ref{SWaveCondRLZ} and \ref{SWaveCondRLF}. However, for the p-wave insulator and superconductor phase transition with Weyl corrections, due to the direct dependence of Eq. (\ref{pwave-Psi}) for the scalar field $\psi$ on $\gamma$, the Weyl correction terms $\gamma$ will appear in the master equation (\ref{NewPWPsiCritical}), this results in the dependence of the critical chemical potential $\mu_{c}$ on the Weyl coupling parameters $\gamma$ in this case, which agrees well to the numerical results. Thus, the Weyl corrections have completely different effect on the critical chemical potential for the s-wave and p-wave insulator/superconductor phase transitions.

On the other hand, the effect of Weyl corrections on the s-wave insulator/superconductor phase transition is reminiscent of that seen for the holographic insulator/superconductor phase transition with $F^{4}$ corrections discussed in Ref. \cite{PJWPRD}, where the Maxwell field strength corrections $F^{4}$ do not influence the s-wave insulator and superconductor phase transition. Thus, it is interesting to note that, for the fixed mass of the scalar field, the critical chemical potential $\mu_{c}$ is independent of the corrections to the Maxwell field, which may be a quite general feature for the s-wave holographic insulator and superconductor model.

For completeness, we still work on Eq. (\ref{SWCriticalPsi}) to understand the dependence of the critical chemical potential on the mass of the scalar field analytically. Defining a trial function $F(z)$ near the
boundary $z=0$ as
\begin{eqnarray}\label{SWPhiFz}
\psi(z)\sim \langle{\cal O}_{i}\rangle z^{\Delta_i}F(z),
\end{eqnarray}
we can obtain the equation of motion for $F(z)$
\begin{eqnarray}\label{NSWFzmotion}
(MF^{\prime})^{\prime}+M\left[\frac{\Delta_i(\Delta_{i}-1)}{z^2}+\frac{\Delta_i}{z}
\left(\frac{f'}{f}-\frac{1}{z}\right)+\frac{1}{z^{4}f}(\mu^{2}z^{2}-m^{2})\right]F=0,
\end{eqnarray}
where we have introduced a new function
\begin{eqnarray}
M(z)=z^{2\Delta_{i}-3}(z^{4}-1).
\end{eqnarray}
The boundary conditions for $F(z)$ are $F(0)=1$ and $F'(0)=0$.

Following the Sturm-Liouville eigenvlaue problem \cite{Gelfand-Fomin}, we deduce the expression which can be used to estimate the minimum eigenvalue of $\mu^2$
\begin{eqnarray}\label{SWeigenvalue}
\mu^{2}=\frac{\int^{1}_{0}M\left(F'^{2}-PF^{2}\right)dz}{\int^{1}_{0}WF^{2}dz},
\end{eqnarray}
with
\begin{eqnarray}
&&P=\frac{\Delta_{i}(\Delta_{i}-1)}{z^{2}}+\frac{\Delta_{i}}{z}\left(\frac{f'}{f}-\frac{1}{z}\right)-\frac{m^{2}}{z^{4}f},\nonumber\\
&&W=\frac{M}{z^{2}f}.
\end{eqnarray}
In the following calculation, we still assume the trial function to
be $F(z)=1-az^{2}$, where $a$ is a constant.

As an example, we will calculate the critical chemical potential $\mu_{c}$ for the case of $i=+$ and one can easily extend the study to the case of $i=-$. From Eq. (\ref{SWeigenvalue}), we obtain
\begin{eqnarray}
\mu^{2}=\frac{\Xi(a,m)}{\Sigma(a,m)},
\end{eqnarray}
with
\begin{eqnarray}
\Xi(a,m)&=&-\frac{8+m^{2}+4\sqrt{4+m^{2}}}{2}\left(\frac{1}{2+\sqrt{4+m^{2}}}-
\frac{2a}{3+\sqrt{4+m^{2}}}+\frac{a^{2}}{4+\sqrt{4+m^{2}}}\right)
\nonumber\\&&
-\frac{4a^{2}}{12+m^{2}+6\sqrt{4+m^{2}}},
\nonumber\\
\Sigma(a,m)&=&-\frac{1}{2(1+\sqrt{4+m^{2}})}+\frac{a}{2+\sqrt{4+m^{2}}}-\frac{a^{2}}{2(3+\sqrt{4+m^{2}})}.
\end{eqnarray}
Hence we can get the minimum eigenvalue of $\mu^{2}$ and the corresponding value of
$a$ for different values of the mass of scalar field, for example, $\mu_{min}^{2}=11.607$ and $a=0.440$ for
$m^{2}L^{2}=0$, $\mu_{min}^{2}=9.843$ and $a=0.423$ for
$m^{2}L^{2}=-1$, $\mu_{min}^{2}=7.936$ and $a=0.401$ for
$m^{2}L^{2}=-2$, $\mu_{min}^{2}=5.753$ and $a=0.371$ for
$m^{2}L^{2}=-3$, and $\mu_{min}^{2}=3.574$ and $a=0.330$ for
$m^{2}L^{2}=-15/4$, which lead to the critical chemical potential
$\mu_{c}=\mu_{min}$ \cite{Cai-Li-Zhang}. In Table
\ref{SWCriticalZheng}, we give the critical chemical potential
$\mu_{c}$ for chosen values of the scalar field. Comparing with numerical results, we find that
the analytic results derived from S-L method agree well with the numerical calculation.

\begin{table}[ht]
\begin{center}
\caption{\label{SWCriticalZheng} The critical chemical potential $\mu_{c}$ for the operator $\langle{\cal O_{+}}\rangle$ obtained by the analytical S-L method and from numerical calculation with chosen various masses of the scalar field for the s-wave holographic insulator and
superconductor model. In order to compare with the results in Ref. \cite{Nishioka-Ryu-Takayanagi}, we also present the critical chemical potential for $m^{2}L^{2}=-15/4$. Note that the Weyl corrections do not have any effect on the critical chemical potential $\mu_{c}$ for the fixed mass of the scalar field.}
\begin{tabular}{c c c c c c}
         \hline \hline
$m^{2}L^{2}$ & 0 & -1 & -2 & -3 & -15/4
        \\
        \hline
~~~~~$Analytical$~~~~~~&~~~~~~$3.407$~~~~~~&~~~~~~$3.137$~~~~~~&~~~~~~$2.817$~~~~~~&~~~~~~$2.399$~~~~~~&~~~~~~$1.890$~~~~~
          \\
~~~~~$Numerical$~~~~~~&~~~~~~$3.404$~~~~~~&~~~~~~$3.135$~~~~~~&~~~~~~$2.815$~~~~~~&~~~~~~$2.396$~~~~~~&~~~~~~$1.888$~~~~~
          \\
        \hline \hline
\end{tabular}
\end{center}
\end{table}

From Table \ref{SWCriticalZheng}, we observe that, with the increase of the mass of scalar field, the critical chemical potential $\mu_{c}$  becomes larger. This property also agrees well with the numerical result \cite{Pan-Wang}. However, the Weyl corrections do not have any effect on the critical chemical potential $\mu_{c}$ for the fixed mass of the scalar field, which can be used to back up the numerical
finding as shown in Figs. \ref{SWaveCondRLZ} and \ref{SWaveCondRLF}.

\subsection{Critical phenomena}

Consider that the condensation of the scalar operator $\langle{\cal
O}_{i}\rangle$ is so small when $\mu\rightarrow\mu_{c}$, we can therefore expand $\phi(z)$ in small
$\langle{\cal O}_{i}\rangle$ as
\begin{eqnarray}\label{SWavePhiChiz}
\phi(z)\sim\mu_{c}+\langle{\cal O}_{i}\rangle\chi(z)+\cdots,
\end{eqnarray}
with the boundary condition $\chi(1)=0$ at the tip. Just as stated for Eq. (\ref{PWPhiChiz}) in the p-wave model, Eq. (\ref{SWavePhiChiz}) is only valid right above the critical point $\mu_{c}$. Using the function defined in Eq. (\ref{PWQz}) and substituting the function (\ref{SWPhiFz}) into Eq. (\ref{SWPhiz}), we can obtain the equation of motion for $\chi(z)$
\begin{eqnarray}\label{SWChi}
(Q\chi')'-2\langle{\cal O}_{i}\rangle\mu_{c}\frac{z^{2\Delta_i-5}(z^{4}-1)F^{2}}{f}=0,
\end{eqnarray}
and its general solution
\begin{eqnarray}\label{SWGSChiz}
&\chi(z)&=\langle{\cal O}\rangle\mu_{c}\zeta(z)
\nonumber\\
&&=\langle{\cal O}\rangle\mu_{c}\left\{c_{1}+\int_{1}^{z}\left[c_{2}+
2\int_{1}^{y}\frac{x^{2\Delta_i-5}(x^{4}-1)F(x)^{2}}{f(x)}dx\right]\frac{1}{Q(y)}dy\right\},
\end{eqnarray}
where $c_{1}$ and $c_{2}$ are the integration constants which can be determined by the
boundary condition of $\chi(z)$.

From the asymptotic behavior in Eq. (\ref{SWInfinity}), we can
expand $\phi$ when $z\rightarrow0$ as
\begin{eqnarray}\label{SWExpPhi}
\phi(z)\simeq\mu-\rho z^{2}\simeq\mu_{c}+\langle{\cal
O}_{i}\rangle[\chi(0)+\chi'(0)z+\frac{1}{2}\chi''(0)z^{2}+\cdots].
\end{eqnarray}
Thus, according to the coefficients of the $z^{0}$ term, we get
\begin{eqnarray}
\mu-\mu_{c}\simeq\langle{\cal O}_{i}\rangle\chi(0),
\end{eqnarray}
which results in
\begin{eqnarray}\label{SWOperatZF}
\langle{\cal
O}_{i}\rangle=\frac{1}{[\mu_{c}\zeta(0)]^{1/2}}(\mu-\mu_{c})^{1/2},
\end{eqnarray}
where $\zeta(0)$ can be determined by Eq. (\ref{SWGSChiz}). For example, fixing $m^{2}L^{2}=-15/4$ and
$\gamma=-0.06$, we can get $\langle{\cal O_{+}}\rangle\approx1.652(\mu-\mu_{c})^{1/2}$ when $a=0.330$, which agrees well with the numerical result given in Fig. \ref{SWaveCondRLZ}. Especially, for the case of $m^{2}L^{2}=-15/4$ and $\gamma=0$ when $a=0.330$, we obtain $\langle{\cal O_{+}}\rangle\approx1.801(\mu-\mu_{c})^{1/2}$, which is in good agreement with the result given in \cite{Nishioka-Ryu-Takayanagi,Cai-Li-Zhang}.

Since the expression (\ref{SWOperatZF}) is valid for all cases considered here, so near the critical point, both of the scalar operators $\langle{\cal O}_{+}\rangle$ and $\langle{\cal O}_{-}\rangle$ satisfy $\langle{\cal O}_{i}\rangle\sim(\mu-\mu_{c})^{1/2}$. This behavior holds for various values of Weyl coupling parameters and masses of the scalar field. The analytic result supports the numerical computation shown in Figs. \ref{SWaveCondRLZ} and \ref{SWaveCondRLF} that the phase transition between the s-wave holographic insulator and superconductor belongs to the second order and the critical exponent of the system takes the mean-field value $1/2$. The Weyl corrections will not influence the result.

Considering the coefficients of $z^{1}$ terms in Eq. (\ref{SWExpPhi}), we point out that $\chi'(0)\rightarrow0$ if $z\rightarrow0$, which is consistent with the following relation by integrating both sides of Eq. (\ref{SWChi})
\begin{eqnarray}\label{SWChiz0}
\left[\frac{\chi'(z)}{z}\right]\bigg|_{z\rightarrow 0}=2\langle{\cal
O}_{i}\rangle\mu_{c}\int_{0}^{1}\frac{z^{2\Delta_i-5}(z^{4}-1)F^{2}}{f}dz.
\end{eqnarray}

Comparing the coefficients of the $z^{2}$ term in Eq. (\ref{SWExpPhi}), we can express $\rho$ as
\begin{eqnarray}
\rho=-\frac{1}{2}\langle{\cal O}_{i}\rangle\chi''(0)=\Gamma(\gamma,m)(\mu-\mu_{c}),
\end{eqnarray}
where $\Gamma(\gamma,m)$ is a function of the Weyl coupling parameter and the scalar field mass
\begin{eqnarray}
\Gamma(\gamma,m)=-\frac{1}{\zeta(0)}\int_{0}^{1}\frac{z^{2\Delta_i-5}(z^{4}-1)F^{2}}{f}dz.
\end{eqnarray}
For the scalar operator $\langle{\cal O}_{+}\rangle$, as an example, fixing $m^2L^2=-15/4$ and $\gamma=-0.06$ when $a=0.330$, we can get $\rho=1.119(\mu-\mu_{c})$, which is in good agreement with the result shown in Fig. \ref{SWaveCondRLZ}. Note that fixing $m^2L^2=-15/4$ and $\gamma=0$ when $a=0.330$, we
can get $\rho=1.330(\mu-\mu_{c})$ for considering the scalar operator $\langle{\cal O}_{+}\rangle$, which is consistent with the result given in \cite{Cai-Li-Zhang}. Here we observed again that the Weyl corrections will not alter the result. Our analytic finding of a linear relation between the charge density and the chemical potential $\rho\sim(\mu-\mu_{c})$ supports the numerical result presented in Figs. \ref{SWaveCondRLZ} and \ref{SWaveCondRLF}.

\section{Conclusions}

We have investigated analytically the condensation and critical phenomena of the phase transition between the holographic insulator and superconductor with Weyl corrections in the probe limit by using the S-L method in order to understand the influences of the $1/N$ or $1/\lambda$ corrections on the holographic dual model in the AdS soliton background. Both in p-wave (the vector field) and s-wave (the scalar field) models, we obtained analytically the critical chemical potentials which are perfectly in agreement with those obtained from numerical computations. We observed that similar to the curvature corrections, in p-wave model, the higher Weyl corrections will make it harder for the holographic insulator/superconductor phase transition to be triggered. However, the story is completely different if we study the s-wave model. In contrast to the effect of curvature corrections, we found for this case that the critical chemical potentials are independent of the Weyl correction terms, which tells us that the Weyl couplings will not affect the properties of the holographic insulator/superconductor phase transition. This behavior is reminiscent of that seen for the holographic insulator and superconductor phase transition model with $F^{4}$ corrections where the Maxwell field strength corrections do not influence the s-wave insulator/superconductor phase transition \cite{PJWPRD}. Thus, we interestingly noted that the corrections to the Maxwell field do not have any effect on the critical chemical potential, which may be a quite general feature for the s-wave holographic insulator/superconductor phase transition.

Furthermore, we discussed analytically the type of phase transition and the relation between the charge density and the chemical potential near the phase transition point. We found that the effect of the Weyl corrections cannot modify the critical phenomena, and found that the holographic insulator/superconductor phase transition belongs to the second order and the critical exponent of the system always takes the mean-field value $1/2$ in both p-wave and s-wave models. The results may be natural since the deviations from the mean-field behaviors do not occur in the superconductors with these higher derivative corrections within the framework of AdS/CFT correspondence \cite{WuCKW,MaCW,MomeniSL,WeylC}. Our analytic results can be used to back up the numerical findings in the holographic insulator and superconductor model with Weyl corrections.

\begin{acknowledgments}

This work was supported by the National Natural Science Foundation of China under Grant Nos. 11275066 and 11175065; the National Basic Research of China under Grant No. 2010CB833004; PCSIRT under Grant No. IRT0964; Hunan Provincial Natural Science Foundation of China under Grant Nos. 12JJ4007 and 11JJ7001; and the Construct Program of the National Key Discipline.

\end{acknowledgments}

\end{document}